\documentstyle[epsf,psbox,12pt]{WORKSHOP}
\newcommand{\SiTh}{{$0.2\pm 0.1\,\mu$m}}
\newcommand{\SiOTh}{{$0.6\pm 0.2\,\mu$m}}
\begin{document}
\pagestyle{empty}

\large
\title{X-ray Measurement of the subpixel structure of\\
 the XMM EPIC MOS CCD}

\author{\\
H.~{\sc Tsunemi} $^{a,b}$,
K.~{\sc Yoshita} $^{a}$,
A.D.~{\sc Short} $^{c}$,
P. J.~{\sc Bennie} $^{c}$,\\
M.J.L.~{\sc Turner} $^{c}$,
A. F.~{\sc Abbey} $^{c}$\\
$^{a}$Department of Earth and Space Science, Graduate School
of Science, Osaka University, \\
1-1 Machikaneyama-cho, Toyonaka, Osaka 560-0043, Japan\\
$^{b}$CREST, Japan Science and Technology Corporation (JST)\\
$^{c}$X-ray Astronomy Group, Department of Physics \&
Astronomy, Leicester University
\\
\\
Accepted in Nucl. Instr. and Methods A, 1999
}

\abst {
We report here the results of a mesh experiment to measure the
subpixel structure of the EPIC MOS CCDs on board the XMM X-ray
observatory.  The pixel size is $40\,\mu$m square while the mesh hole
spacing is $48\,\mu$m, a combination quite different from our standard
mesh experiment.  We have verified that this combination functions
properly and have analyzed the CCD structure with sub-pixel
resolution.

The EPIC MOS CCD has an open electrode structure to improve detection
efficiency at low energies.  We obtained the distribution of various
grades of X-ray events inside the pixel.  A horizontally split
two-pixel event is generated near the channel stop which forms a
straight vertical pixel boundary whereas a vertically split two-pixel
event is generated where the potential due to the thinned gate
structure forms a wavy horizontal pixel boundary.  Therefore, the
effective pixel shape is not a square but is distorted.

The distribution of X-ray events clearly shows that the two etched
regions in each pixel, separated by the bridging finger of the
enlarged (open) electrode.  We measured the difference in X-ray
transmission between the conventional and open regions of the pixel
using O-K and Cu-L X-ray emission lines, and found it to be consistent
with an electrode thickness comprising \SiTh~of $Si$
and \SiOTh~of $SiO_2$.\\
}

\kword{
charge-coupled-device, mesh experiment, open electrode, subpixel resolution
}

\maketitle
\thispagestyle{empty}

\section{Introduction}

The Charge-Coupled Device (CCD) is now widely used in X-ray photon
counting detector applications, and particularly in X-ray
astronomy~\cite{frazer89}.  CCDs consist of many small pixels each of
which functions as an independent detector.  When an X-ray photon is
absorbed in a CCD, it produces a finite charge cloud.  When the
photo-absorption occurs in the depletion layer, the entire charge
cloud is collected and read out as a signal, but a total charge is
often shared between several adjacent pixels each of which has a
signal greater than the detection threshold.  In this way X-ray events
fall into different grades; single-pixel events, vertically split
two-pixel events, horizontally split two-pixel events, 3--4 pixel split 
events etc.  The type of
event observed depends principally upon the proximity of the X-ray's
absorption to a pixel boundary, as well as on its energy and its depth.
In this sense it is the real pixel boundary as defined by the
potentials within the device, and not simply the geometrical
arrangement of electrodes which is important. The total charge
(proportional to the X-ray energy) can be measured by summing the
signals from all pixels associated with the event.

The response function of a CCD can be divided into three parts: the
gate structure transmission, the absorption efficiency in the
depletion region, and the charge spreading after the photo-electric
absorption. We usually calibrate CCDs by uniformly irradiating them
with X-rays at a number of discreet energies.  In this way, X-rays are
incident with equal probability upon all regions of the pixel, and the
calibration data contain effects which cannot be measured
separately. However, since the thickness of the gate and the depletion
region vary within each pixel, the detection efficiency must also
vary. To obtain a realistic response function of the CCD for the data
analysis, it is therefore necessary to extract the response from
different regions within the pixel.

Recently, a new technique has been introduced to obtain the X-ray
response of the CCD with sub-pixel resolution~\cite{tsunemi97}.  The
technique consists of a metal mesh positioned just above the CCD and a
parallel beam of X-rays. The holes in the mesh are smaller than the
CCD pixel size and have periodic spacing. This technique enables us to
restrict the incident X-ray position with sub-pixel resolution.  The
sub-pixel structures of the various types of the CCD (ASCA SIS, AXAF
ACIS etc.) have been measured using this mesh
technique~\cite{yoshita98,pivovaroff98}.  In this paper, we report on
a measurement of the EPIC MOS (Metal Oxide Semiconductor) \lq open
electrode\rq~CCD developed for the XMM observatory.

\section{EPIC MOS CCDs}

The EPIC cameras~\cite{turner98} are focal plane imaging spectrometers
developed for XMM~\cite{turner99}. There are three cameras, situated
at the foci of the three mirror modules, and all carry silicon CCD
detectors. One of the cameras utilizes PN technology CCDs and has been
developed by the Max Plank Institute. The other two cameras carry MOS
CCDs (EEV CCD22) and have been developed by the X-ray Astronomy Group
at Leicester University~\cite{alex98}. The EEV CCD22 is a frame
transfer, front-illuminated device. The image section consists of a
600$\times$600 arrays of 40\,$\mu$m square pixel.

\begin{figure}
\begin{center}
\psbox[xsize=0.32#1,ysize=0.32#1]{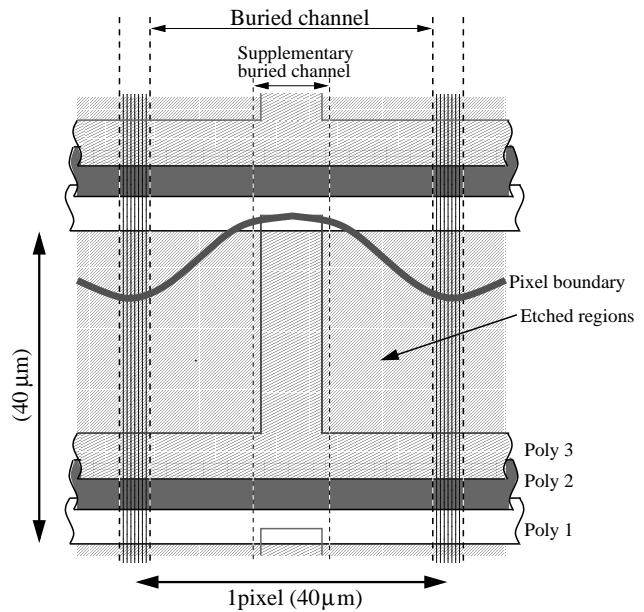} 
\end{center}
\caption{Schematic view of the gate structure~[7]
of the CCD we employed.  
There are three gates: one of them is partly thinned in
order to improve the detection efficiency at low energy.  The wavy
line denotes the horizontal pixel boundary.}
\label{gate_structure}
\end{figure}

The CCD22 is a three phase device, and the electrodes, or gates (poly
l, 2 and 3) are shown schematically in
Fig.~\ref{gate_structure}~\cite{alex98}.  In order to obtain a useful
X-ray detection efficiency at low energies, one of the gates, poly-3,
has been enlarged and two holes have been left, through this enlarged
gate to the native oxide layer. These two holes are separated by a
central electrode \lq finger\rq.  A P$^+$ dopant is implanted in the
etched areas, which pins the surface potential to the substrate
potential.

\section{Experimental Setup}

The experiment was performed in a CCD test facility at Leicester
University. The principle of the mesh experiment is described in
Tsunemi {\it et al.}~\cite{tsunemi97} There are normally two types of
mesh experiment: a single-pitch mesh experiment and a multi-pitch mesh
experiment~\cite{tsunemi98}. In the single-pitch mesh experiment, we
employ a mesh whose hole spacing is equal to that of the CCD pixel
size, while in the multi-pitch mesh experiment, we employ a mesh whose
hole spacing is a multiple of the CCD pixel size. For this experiment
we employed a copper mesh with a thickness of 10\,$\mu$m. It has holes
of 3.4\,$\mu$m diameter, periodically spaced at 48\,$\mu$m
intervals. This mesh was originally designed in order to perform the
multi-pitch mesh experiment using CCDs with $12\,\mu$m square
pixels~\cite{hiraga98}. A novel feature of this experiment was
therefore that the mesh hole spacing is not an integer multiple of the
CCD pixel size. This makes the analysis slightly more complicated. The
mesh was placed approximately 2\,mm above the CCD and was rotated
slightly with respect to the CCD edges.

\begin{figure}
\begin{center}
\psbox[r,xsize=0.35#1,ysize=0.35#1]{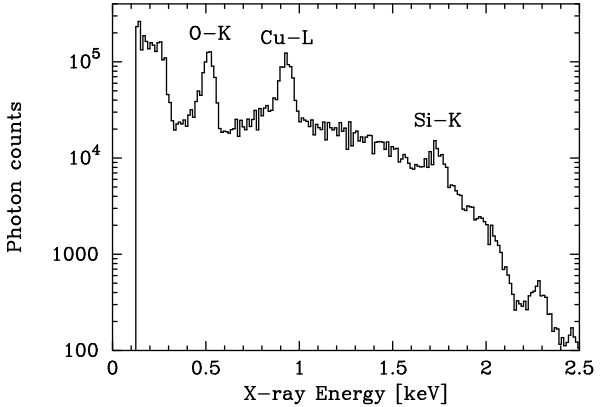}
\end{center}
\caption{X-ray spectrum obtained with single-pixel events.  There are
three characteristic X-ray emissions: O-K, Cu-L and Si-K.  We employed
O-K and Cu-L for the analysis.}
\label{single_spectrum}
\end{figure}

The X-ray source was approximately 3\,m from the CCD and several
fluorescence targets were used, generating characteristic X-rays as
well as a bremsstrahlung spectrum. We mainly selected the emission
lines for analysis. Figure~\ref{single_spectrum} shows a typical X-ray
spectrum obtained by selecting only single-pixel events. It includes
emission lines of O-K (0.52\,keV), Cu-L (0.93\,keV) and Si-K
(1.74\,keV). The energy range below 0.3\,keV is enhanced by C-K
(0.28\,keV). In our experiment, we only used the O-K and Cu-L emission
lines and we placed a pinhole of 5\,mm diameter in front of the
X-ray generator to restrict the beam divergence, but the effective
mesh hole shadow on the CCD was still estimated to be about 7\,$\mu$m. 
The X-ray generator current was restricted so that X-ray photons did
not heavily pile up on the CCD.

The CCD operating conditions were identical to those employed on the
XMM satellite. The chip was cooled to $-100^\circ$C using liquid
nitrogen and was driven using duplicate flight electronics.

\section{Mesh experiment}

In the mesh experiment, the incident X-ray position on the CCD is
restricted by the mesh hole size which is smaller than the CCD pixel
size. Equation~(\ref{relation}) shows the relation between the CCD
coordinate, {\bf X}, and the mesh coordinate, {\bf x}.
\begin{eqnarray}
{\bf X}\!\! &=&\!\! M {\bf x} + {\bf X_{off}} \nonumber \\
\!\!&=&\!\! m\left(\begin{array}{cc} \!\!1+a & \!0\!\! \\ \!\!0 & \!1+b\!\! \end{array} \right)\!
\left( \begin{array}{cc} 
\!\!\cos\theta & \!-\sin\theta\!\!  \\
\!\!\sin\theta & \!\cos\theta\!\!  \end{array}  \right) 
{\bf x} \nonumber \\
& &\!\! + {\bf X_{off}} \label{relation}
\end{eqnarray}
where $m$ is the mesh experiment multiplier, $a$ and $b$ are
expansion coefficients along the x and y axes, $\theta$ is the tilt angle
and ${\bf X_{off}}$ is an offset.

Therefore, in order to reconstruct the sub-pixel response of the CCD,
it is first necessary to determine these parameters for the
experimental alignment between the mesh and the CCD with sub-pixel
precision. The standard mesh experiment is performed with $m$ an
integer~\cite{tsunemi97,yoshita98,pivovaroff98,tsunemi98,hiraga98} and
the raw data clearly shows a moire pattern from which approximate
values for the parameters may be calculated. Then we can easily
determine the mutual alignment between the mesh and the CCD with
sufficient precision to reconstruct the sub-pixel response. The
detailed data reduction method is described
in~\cite{tsunemi98,tsunemi99}.

\begin{figure}
\begin{center}
\psbox[xsize=0.5#1,ysize=0.5#1]{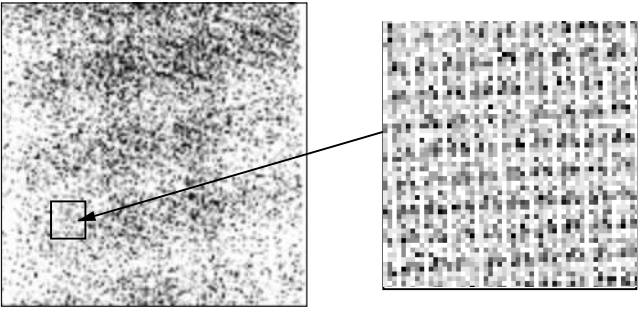}
\end{center}
\caption{Raw image of single-pixel events shows a moire pattern.  We
see a periodic structure generated by the resonance between the CCD
pixel spacing and the mesh hole spacing.}
\label{g0image}
\end{figure}

In this experiment, $m$ has a non-integer value of
1.2. Figure~\ref{g0image} shows a raw image of Cu-L single-pixel
events. The intensity is not uniform, but shows a resonance between
the CCD pixel size and the mesh hole spacing. Assuming that
single-pixel events occur when X-rays are incident upon the inner part
of the pixel, we calculate a parameter $Dist$, the distance between a
pixel giving single-pixel events and its nearest associated hole
position.  This parameter is defined in
eq.(\ref{dist})\cite{tsunemi99c}.

\begin{eqnarray}
Dist(a, b, \theta,  {\bf X_{off}} ) & = &\sum_{i}^{pixel} ({\bf X_{hole}} - {\bf X_i})^2 N_i\hspace{3mm} ,  \nonumber\\
{\bf X_{hole}} & = &M~{\bf x_{hole}} + {\bf X_{off}} \label{dist}
\end{eqnarray}
where ${\bf X_i}$ is the position of the CCD of the {\it i}\,th pixel,
$N_i$ is the number of single-pixel events detected in the {\it i}\,th
pixel, and ${\bf x_{hole}}$ is the hole position nearest to the {\it
i}\,th pixel.  $Dist$ is a function of $a$, $b$, $\theta$ and ${\bf
X_{off}}$ and the true values of these parameters may be found by
minimizing $Dist$.  In this way, we determined the best fit values as
$a = 0.8\times 10^{-3}$, $b = 0.9\times 10^{-3}$, $\theta =
0.09^{\circ}$ and ${\bf X_{off}} = (0.81, 0.33)$.

\section{Restored Image}

In the mesh experiment, there are two ways to restore the image: one
is restoration onto the CCD coordinate system, and the other is
restoration onto the mesh coordinate system~\cite{tsunemi99}. In this,
non-standard experiment, we restored the image onto the mesh
coordinate system. The restored image is a \lq representative
unit\rq~(RU) of $48\,\mu$m square which is a convolution of the
effective holes and the CCD pixel structure ($40\,\mu$m square).

\begin{figure}
\begin{center}
\psbox[xsize=0.5#1,ysize=0.5#1]{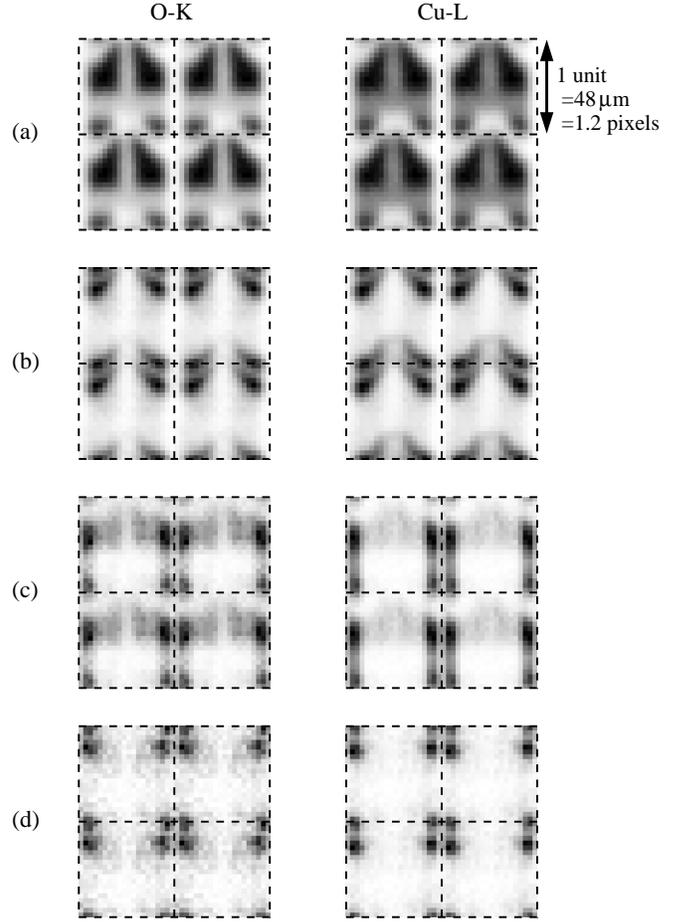}
\end{center}
\caption{2$\times$2 representative units (on the mesh coordinate) for (a)single-pixel events, (b)vertically split two-pixel events, (c)horizontally
split two-pixel events and (d)corner events. Restored images for O-K
are in the left column while those for Cu-L are in the right column.}
\label{unit_mesh_hole}
\end{figure}

Figure~\ref{unit_mesh_hole} shows examples of $2\times2$ RUs for
various event grades.  We set the split threshold level to be 63\,eV
throughout our analysis. The RU for single-pixel events has four
isolated parts. In the RU for vertically split two-pixel events, we clearly
see four wavy structures near the top and bottom of each pixel. The
structures near the top correspond to two-pixel split events in which
the greater charge is in the lower pixel, while those near the bottom
corresponds to two-pixel split events in which the greater charge is
in the upper pixel. Therefore, the gap between these regions,
corresponds to the pixel boundary.  We should note that the horizontal
pixel boundary is not a straight line but is curved.  However, the
pixel boundary in the vertical direction is a straight line which is
seen in the RU for horizontally split two-pixel events. The 3--4 pixel
split events appear near the pixel corners as one would expect.

A pixel boundary is a region where the electric potential forms a
barrier to electrons during integration.  The vertical boundaries are
formed by the channel stops, and the gates form the horizontal
boundaries.  The poly-3 gate has been partially removed in order to
increase the low energy detection efficiency, leaving a central finger
in order to transfer the charge to the adjacent pixel.  The schematic
view in Fig.~\ref{gate_structure} shows the predicted horizontal pixel
boundary as a thick, wavy line.  The pixel boundaries in conventional
CCD chips are straight lines in both the horizontal and vertical
directions, the wavy boundary being a feature of the \lq open\rq~gate
structure.

\begin{figure}
\begin{center}
\psbox[xsize=0.5#1,ysize=0.5#1]{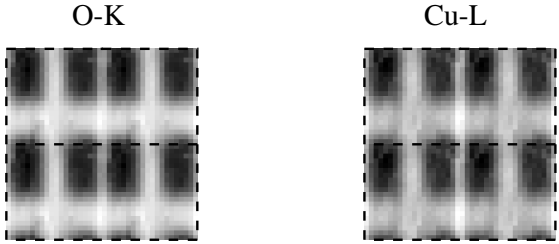}
\end{center}
\caption{2$\times$2 representative pixel (on the CCD coordinate) arrays for all events.
Dashed lines represent the $40\,\mu$m square array.}
\label{unit_gall}
\end{figure}

We reconstructed the CCD pixel image using all X-ray events, taking
into account the wavy boundary.  The reconstructed image is a \lq
representative pixel\rq~(RP) of $40\,\mu$m square.
Figure~\ref{unit_gall} shows the CCD responsivity of $2\times2$ RPs.
We clearly see that there are two regions in the pixel where the X-ray
responsivity is enhanced particularly for low energy X-rays.  These
enhanced regions coincide with the open sections of the pixel.

Due to the relatively large effective diameter of the mesh holes, it
was not possible to accurately measure the shape of the thinned gate
area during this experiment.  From other experiments, scanning electron
microscope (SEM) images, and information gained during device fabrication
however, the open area is known to be about 40\,\% of the total pixel
area.  Therefore, we simply assumed that 40\,\% of each pixel has an
enhanced responsivity while the remainder is covered by a uniform
conventional gate structure.  This assumption does not realistically
represent the gate overlaps, the difference in thickness between the
different gates or the channel stops.  These fine structures will be
studied in detail in the next experiment, which will utilize a mesh
with a much finer effective mesh hole size.

Based on our simple assumption, we measured the difference in the
electrode absorption between the thinned gate region and the rest of
the pixel, in terms of an effective absorber thickness.  For this we
used O-K and Cu-L data for which we had sufficient statistics.  Since
these two lines are respectively below and above the O-K edge, we are
able to determine the differences in thickness between the thinned
gate area and the rest of the pixel, for $Si$ and $SiO_2$
independently. They are \SiTh~for $Si$ and \SiOTh~for $SiO_2$. We
should note that the errors in these two results are strongly
correlated with each other.

\section{Discussion}

X-rays detected by a photon counting CCD form events which fall into
several grades (single-pixel, two-pixel etc.).  The grade of event
depends principally upon the position within the CCD pixel at which
the X-ray photon is absorbed. The pixel boundaries of conventional
CCDs form a simple
square~\cite{tsunemi97,yoshita98,pivovaroff98,hiraga98,tsunemi99c},
the vertical pixel boundary being formed by the channel stops which
are straight lines, and the horizontal pixel boundaries being formed
by clocking gates which are also straight lines.

The EEV CCD22, having an open electrode structure, exhibits quite
different boundaries.  The vertical pixel boundary is still a straight
line since the channel stops are straight, but the horizontal pixel
boundaries are wavy due to the open gate structure.  This is because
the potential in some parts of the open region is determined by the
electrode of the neighboring pixel, so that the potential barrier
separating the pixels is located within the open area and not under
the electrodes.
 
Since the horizontal pixel boundary runs through the open electrode
region, some of the X-ray photons incident upon this open region form
charge clouds which are collected in the neighboring pixel, and others
give rise to vertically split two-pixel events rather than
single-pixel events as seen in Figure~\ref{unit_mesh_hole}.  Since the
X-ray photon energy is measured by summing the charge from all pixels
comprising the event, a single-pixel event usually shows better energy
resolution than split-pixel events.  Table~\ref{branching_ratio} shows
the branching ratio i.e. the fraction of X-ray events forming each
grade of events.  These values are obtained by eliminating the pile-up
effects.

\begin{table}
\caption{Branching ratio of the X-ray event grades}
\label{branching_ratio}
\begin{tabular}{ccc} \hline
                        	  & 	O-K 	&     Cu-L	\\ \hline
Energy (keV)           		  &     0.52	&     0.93	\\ \hline
Single-pixel (\%)		  &     81.3	&     79.0	\\
Vertically split two-pixel (\%)	  &	15.3	&     16.2	\\
Horizontally split two-pixel (\%) &     3.0	&     4.0	\\
3--4 pixel (\%)			  &	0.4	&     0.8	\\
\hline
\end{tabular}
\end{table}

We also observe that vertically split two-pixel events are detected
with a greater efficiency than horizontally split two-pixel events.
This is consistent with them being generated in the open region rather
than under the electrodes.  The ratio between the horizontally split
two-pixel events and the vertically split two-pixel events depends
partly on the charge cloud shape and size, and has previously been
used as a diagnostic~\cite{hiraga98,tsunemi99b,yoshita99}.  However, in this case, the difference in
detection efficiency for the two grades of event significantly alters
the branching ratio, making it difficult to evaluate the charge cloud
shape.

We restored the CCD pixel image using all the X-ray events and found
that there are two regions in each pixel where the detection
efficiency is enhanced.  There are four regions where single-pixel
events are generated, two of which are separated by the regions where
the vertically split two-pixel events are generated.  These regions
are located in the open electrode area, and this is confirmed by
increased detection efficiency.
 
The XMM MOS CCD has a relatively large pixel size ($40\,\mu$m square).
This is significantly larger than other devices to which this
technique has been applied including the ASCA SIS~\cite{yoshita98}
($27\,\mu$m) and AXAF ACIS~\cite{pivovaroff98} ($24\,\mu$m). However,
the majority of the pixel is occupied by the third phase, much of
which is open (i.e. is not covered by polysilicon electrode) in order
to increase the detection efficiency at low energies.  The scale of
the gate structures and overlaps themselves are of the order of one
$\mu$m.  In order to obtain the precise response function of the CCD
in future experiments it will therefore be necessary to use a mesh
with a smaller hole size and a less divergent X-ray beam.

\section{Conclusion}

We performed a mesh experiment using a CCD with an open gate structure
developed for the XMM satellite.  The mesh hole spacing was 1.2 times
greater than the CCD pixel spacing which is a different configuration
from the standard mesh experiment.  We confirmed that this
configuration functions properly although a completely different moire
pattern results.  We employed a distance measure to determine the
parameter values for the mesh experiment.  In this way, we have
succeeded in restoring an image with which we can see where the X-rays
are absorbed within the CCD pixel.

The distribution of the various X-ray event grades are quite different
from those obtained using more conventional CCDs.  The vertical pixel
boundary is represented by a straight line which is formed by the
channel stop.  However, the horizontal pixel boundary is represented
by a wavy line between the poly-l and the poly-3 electrodes.  The
horizontally split two-pixel events are generated in the region near
the channel stop whereas the vertically split two-pixel events are
generated in the open electrode region where the detection efficiency
is enhanced. The effective pixel shape is not a $40\mu$m square but a
distorted shape with an area equal to a $40\mu$m square.

Vertically split two-pixel events are much more numerous than
horizontally split two-pixel events, particularly at low energies.  We
confirmed that this is because they are formed by X-ray absorption in
the open gate region where the detection efficiency is greatly
enhanced.  By comparing the detection efficiency from the open part of
the pixel with an average value for the rest of the pixel, an
equivalent absorber thickness of \SiTh~for $Si$ and \SiOTh~for $SiO_2$.
In order to better resolve the pixel structure for the CCD data
analysis, the next experiment will utilize a mesh with smaller holes
and a less divergent X-ray beam.

\section{Acknowledgements}

We wish to express our thanks to all the member of the X-ray Astronomy
Group in Leicester University for their assistance and comment.  KY is
partly supported by JSPS Research Fellowship for Young Scientists,
Japan.


\begin{thebibliography}{99}
\thispagestyle{empty}

\bibitem{frazer89} G.~W. Frazer: X-ray Detectors in Astronomy
(Cambridge University Press, Cambridge, 1989) p.~208.

\bibitem{tsunemi97} H.~Tsunemi, K.~Yoshita, and S.~Kitamoto,
{\it Jpn J. Appl. Physic}, {\bf 36}, (1997) 2906.

\bibitem{yoshita98} K.~Yoshita, H.~Tsunemi, K.~C.~Gendreau,
G.~Pennington and M.~W.~Bautz, {\it IEEE Trans. Nucl. Sci.}, {\bf 45},
(1998) 915.

\bibitem{pivovaroff98} M.~J.~Pivovaroff, S.~Jones, M.~Bautz,
S.~Kissel, G.~Prigozhin, G.~Ricker, H.~Tsunemi and E.~Miyata, {\it
IEEE Trans. Nucl. Sci.}, {\bf 45}, (1998) 164.

\bibitem{turner98} K.~O. Mason, G.~Bignami, A.~C. Brinkman,
A. Peacock, {\it Advances in Space Research}, {\bf 16}, (1995) 41.

\bibitem{turner99} D.~H. Lumb, H.~Eggel, R.~Laine, A.~Peacock,
{\it Proc. SPIE}, {\bf 2808}, (1996) 326.

\bibitem{alex98} A.~D. Short, A. Keay, M.~J.~L. Turner,
{\it Proc. SPIE}, {\bf 3445}, (1998) 13.

\bibitem{tsunemi98} H.~Tsunemi, J.~Hiraga, K.~Yoshita and S.~Kitamoto,
{\it Jpn J. Appl. Physic}, {\bf 37}, (1998) 2734.

\bibitem{hiraga98} J.~Hiraga, H.~Tsunemi, K.~Yoshita, E.~Miyata and
M.~Ohtani: {\it Jpn. J. Appl. Phys.}, {\bf 37} (1998) 4627.

\bibitem{tsunemi99} H.~Tsunemi, J.~Hiraga, K.~Yoshita and E.~Miyata,
{\it Jpn. J. Appl. Phys.}, {\bf 38} (1999) 2953.

\bibitem{tsunemi99b} H.~Tsunemi, J.~Hiraga, K.~Yoshita, and
K.~Hayashida, {\it Nucl. Instr. and Meth}, {\bf A421} (1999) 90.

\bibitem{tsunemi99c} H.~Tsunemi, J.~Hiraga, K.~Mori, K.~Yoshita and E.~Miyata,
{\it Nucl. Instr. and Meth}, (1999) in press.

\bibitem{yoshita99} K.~Yoshita, H.~Tsunemi, K.~C.~Gendreau,
and M.~W.~Bautz, {\it IEEE trans. NS}, {\bf 46} (1999) 100.

\end{thebibliography}
\end{document}